\documentclass[12pt,preprint]{aastex}

\shorttitle{Iron in Cas A} \shortauthors{Hwang \& Laming}

\begin{document}
\title{Where was the Iron Synthesized in Cassiopeia A?}


\author{Una Hwang\altaffilmark{1} \& J. Martin Laming\altaffilmark{2}}


\altaffiltext{1}{Goddard Space Flight Center and University of Maryland\\
\email{hwang@orfeo.gsfc.nasa.gov}}
\altaffiltext{2}{Code 7674L, Naval Research Laboratory, Washington DC 20375\\
\email{jlaming@ssd5.nrl.navy.mil}}

\begin{abstract}
We investigate the properties of Fe-rich knots on the east limb of the
Cassiopeia A supernova remnant using observations with Chandra/ACIS
and analysis methods developed in a companion paper.  We use the
fitted ionization age and electron temperature of the knots to
constrain the ejecta density profile and the Lagrangian mass
coordinates of the knots.  Fe-rich knots which also have strong
emission from Si, S, Ar, and Ca are clustered around mass coordinates
$q\simeq 0.35-0.4$ in the shocked ejecta; for ejecta mass $2M_{\sun}$,
this places the knots $0.7-0.8 M_{\sun}$ out from the center (or
$2-2.1 M_{\sun}$, allowing for a $1.3 M_{\sun}$ compact object). We
also find an Fe clump that is evidently devoid of line emission from
lower mass elements, as would be expected if it were the product of
$\alpha$-rich freeze out; the mass coordinate of this clump is similar
to those of the other Fe knots.
\end{abstract}


\section{Introduction}
Cassiopeia A is the only supernova remnant for which decays of the
important radioactive nucleus $^{44}$Ti have been confirmed by
observation of $\gamma $ ray line emission from both of its decay
products, $^{44}$Sc and $^{44}$Ca
\citep{iyudin94,iyudin97,vink01,vink03}. The $^{44}$Ti is almost
uniquely produced in the explosive Si burning condition known as
$\alpha$-rich freeze out, wherein Si burns at high temperature and
relatively low density.  It takes place in the innermost ejecta and is
highly sensitive to the explosion details and the position of the mass
cut for the formation of a neutron star or black hole.  The location
of $\alpha$-rich freeze out is thus of great interest and importance
to theoretical progress on the mechanisms by which core-collapse
supernovae explode, but it is technically very difficult to directly
image the hard X-rays or $\gamma$-rays from the $^{44}$Ti decay at the
necessary spatial resolution.  An attractive alternative approach,
recently proposed by \citet{silver01}, is to search for Doppler
structure in the $^{44}$Sc emission lines at 67.9 and 78.4 keV using a
very high spectral resolution calorimeter, but this also must await
developement of the required instrumentation.

A different approach explored in this paper depends on the
nucleosynthesis signature of $\alpha$-rich freeze out.  The usual
explosive Si burning produces mainly $^{56}$Fe (originally as
$^{56}$Ni, which decays radioactively), along with Si, S, Ar, and Ca.
Because $\alpha$-rich freeze out happens very fast at low density and
high temperature, the burning products are almost exclusively
$^{56}$Fe, apart from $^{44}$Ti and a small amount of $^4$He. The
principal reason for this difference is that the low density
suppresses the $3\alpha\rightarrow ^{12}$C reaction, to shift the
nuclear statistical equilibrium (or quasi-equilibrium) to heavier
nuclei around the iron peak \citep{woosley91,arnett96,the98}.  Lighter
nuclei, including $^{44}$Ti, are formed in small amounts towards the
end of the freeze out by $\alpha$ captures. The X-ray spectrum of a
freeze out region should thus show emission lines only from Fe.  It is
our purpose in this paper to locate such Fe-rich region(s) in the
extant Chandra observations of Cas A, and to characterize them in
terms of their temperatures and ionization ages.  By using a model for
the supernova remnant hydrodynamics, we then place limits on the
Lagrangian mass coordinates where such burning took place.

\section{The Structure of Cassiopeia A}

Early Chandra X-ray observations showed that the Fe emission in Cas A
is associated with ejecta and is dominant to the east, in the faint
region outside the bright ejecta shell.  Since the Fe originates in
the innermost ejecta, this suggests a large-scale turbulent overturn
of the ejecta layers, either during or since the explosion
\citep{hughes00}.  In the northwest, the Fe emission coexists
with emission from Si and other intermediate mass elements, at least
in projection, but Doppler measurements made with XMM-Newton
\citep{willingale02} indicate larger line-of-sight velocities for Fe
that imply that the Fe emitting region is actually kinematically and
spatially distinct from that of other elements such as Si.  Overturn
of the X-ray emitting Fe layers thus appears to have occured in the
northwest as well as in the southeast, except that it is not viewed in
a favorable orientation.  Turbulent overturns are also inferred for
the optically emitting S and Ar ejecta \citep{fesen01}.

Typical X-ray surface brightness enhancements for the Fe features are
only a factor of a few above their immediate surroundings.  The
Fe-rich knots will be shown here (also see Laming \& Hwang 2003) to
have higher characteristic ionization ages than the O/Si knots by
factors of a few to ten,\footnote{The ionization age $n_et$ is defined
as the product of electron density and the time since the plasma was
shock-heated.}  implying either a correspondingly higher density or an
earlier shock time.  Because knots and clouds with large density
enhancements (or deficits) are subject to a number of hydrodynamical
instabilities that act to destroy the knots within a few shock
crossing timescales
\citep{klein94,klein03,wang01,poludnenko01,mckee75}, such knots are
not expected to survive longer than several tens of years after
passage of the reverse shock.  We believe that early shock times and
very modest (or no) density enhancements are the most consistent
explanation for the Fe knots that have survived to be observed today,
and we make this assumption in what follows in this paper.  We can
thus account for the observed spectral characteristics of the knots,
while allowing for their low surface brightness enhancements, and
making it plausible for these shocked knots to exist at the present
day.  The low surface brightness enhancement that is observed for the
Fe knots could be effected, not by increased density, but by the high
emissivity of a Fe-rich plasma compared to a solar-abundance plasma at
the same temperature and ionization age.

The survival of knotty ejecta in Cas A is supported by the
hydrodynamical simulations of \citet{kifonidis00,kifonidis03}, who
have modelled core collapse explosions in two dimensions with the aim
of following the propagation of metal clumps within the ejecta.  For a
Type II explosion of a 15 $M_{\sun}$ progenitor, they find that
Rayleigh-Taylor instabilities at the interfaces between the Ni+Si and
O layers and the O+C and He layers produce clumps of widely varying
element compositions.  As the explosion proceeds, however, the blast
wave decelerates in the dense helium layer and generates a reverse
shock that shreds and mixes the clumps.  By a time 20,000 s after
bounce, almost all the metals are completely mixed throughout the
inner 3.4 $M_{\sun}$. The sole exceptions are those species like
$^{44}$Ti that are only synthesized in the innermost layers, and which
remain correspondingly more localized. By contrast, a 15 $M_{\sun}$
progenitor that loses its outer envelope to a stellar wind before
exploding as a Type Ib supernova has a much weaker reverse shock.
This reduces the effect of the mixing instabilities so that the metal
clumps that are formed may survive and propagate further out into the
ejecta.  At time 3,000 s after bounce, $^{28}$Si, $^{44}$Ti and
$^{56}$Ni exist throughout the inner 2 $M_{\sun}$ (of an explosion
progenitor of 5.1 $M_{\sun}$).

There is little doubt that the explosion of Cas A was of Type Ib so
that, based on the above, ejecta knots should have initially survived
the mixing instabilities that operated during the explosion.  For
example, \citet{lamhwa} give various arguments to suggest that the
progenitor's main sequence mass was 20-25 M$_\odot$ (at the lower end
of the range of Wolf-Rayet masses), and that the progenitor possibly
underwent a short phase as a WN star before exploding at 3-4
M$_\odot$.  Similar inferences have been made based on studies of the
optically emitting fast-moving knots \citep[and references
therein]{fesen88, fesen01}.  Indeed, there is substantial evidence to
support a circumstellar environment for Cas A.  Slow-moving, N-rich
optically emitting knots are generally accepted to be relics of the
progenitor's mass loss, and a wind circumstellar density profile is
far more viable than a constant density environment in the light of
such observations as the relative radii of the reverse and forward
shocks \citep{gotthelf01}, the X-ray expansion rate \citep{delaney03},
and also in terms of the implied ejecta mass and the production of
nonthermal X-ray bremsstrahlung emission by electrons accelerated by
lower hybrid waves \citep{laming01a,laming01b,vink03}.

In a companion paper \citep[Paper I]{lamhwa}, we use a new approach to
make a quantitative analysis of the X-ray emitting ejecta of Cas A,
specifically with regard to asymmetries in the ejecta and explosion.
The method of \citet{lamhwa} makes the assumptions that Cas A is
propagating into a circumstellar environment (i.e., the dependence of
the density $\rho$ on the radius $r$ is $\rho \propto r^{-2}$) and
that the spectra of individual knots can be modelled with a single
temperature and ionization age.  Arguments have just been presented
for a circumstellar environment for Cas A.  A single temperature and
ionization age may plausibly be used to model the spectra of the knots
because the knots have angular scales of $\sim 1''$ in the Chandra
images, corresponding to $\sim 5\times 10^{16}$ cm (at a distance of
3.4 kpc).  A 1000 km/s reverse shock can traverse such a knot in just
20 years---a time period much shorter than the timescale for the
dynamical evolution of the remnant.

The model of \citet{lamhwa} extends the analytical hydrodynamical
description of remnant evolution of \citet{truelove99} to a $\rho
\propto r^{-2}$ circumstellar environment, and includes the basic
relevant atomic and plasma physical processes: the time-dependent
ionization balance in shock-heated gas, Coulomb heating of the
electrons by the ions (i.e., no additional collective plasma heating
is assumed), and radiative and adiabatic expansion losses.  The ejecta
themselves are assumed to have a power-law density distribution with a
uniform density core.  The shocked ejecta mass is taken to be 2
M$_\odot$, whereas the explosion energy and ambient density are
adjusted.  These models are used to predict the
time-dependent behavior of the temperature and ionization age of the
knots, which are then compared to the temperatures and ionization ages
inferred for the ejecta knots from their Chandra X-ray spectra.  This
approach allows us to deduce the slope of the ejecta envelope in each
of the azimuthal directions we examine;
by forcing the models to share a common core ejecta density and fixing
the forward shock radius in the models to match the observed average
X-ray radius, we may interpret the inferred variations in the ejecta
envelope slope as due to variations in the locally deposited explosion
energy.  The variations we deduce are modest, at about a factor of two
in energy, with shallower slopes (and higher inferred explosion
energy) at the base of the ejecta jet in the northeast than in the
east or northwest.  A factor of two asymmetry is at the low end
predicted by models, but at present, it is not possible on the basis
of ejecta density and energy asymmetries alone to distinguish between
the two basic mechanisms to produce explosion asymmetries, namely
rotation of the progenitor or a jet-induced explosion
\citep{fryer01,fryer03,khokhlov99}.  Asymmetries of a comparable
magnitude may also be produced by the instability that can arise in
the standing accreting shocks of core-collapse supernovae as the
injection of vorticity drives the rapid growth of turbulence behind
the shock \citep{blondin02}.

\section{Data Reduction and Analysis}

For the imaging analysis and most of the spectral analysis, we use the
same January 2000 Chandra observation of Cas A with the Advanced CCD
Imaging Spectrometer (ACIS) that is used and described by
\citet{lamhwa}.  For one interesting and faint region, however, we
also use data from a second epoch observation of Cas A taken two years
after the first, in January 2002.  Both observations are of 50 ks
duration and were taken with the same instrument in the same
configuration.  The spectra from the two epochs are fitted jointly,
rather than being added into a single spectrum, because of the
time-varying response of the ACIS detector.  These effects include
time-dependent changes in the gain and resolution caused by
charge-transfer inefficiency, and in particular, the temporally
increasing absorption of soft X-rays (with energies below about 1 keV)
by contaminants that build up on the detector surface.  This detector
absorption is modelled by ACISABS in the XSPEC X-ray spectral fitting
package, with the number of days elapsed from launch to the
observation date specified (198 days for the 2000 observation and 929
days for the 2002 observation).

\subsection{Identification of Fe-rich Regions}

In searching for regions dominated by Fe ejecta, we are aided by the
combined spectral and imaging capabilities of the Chandra Observatory.
X-ray images of Cas A in its prominent Fe K blend (energy $\sim$6.7
keV) show that this emission is distributed in three primary
regions---to the southeast, northwest, and west
\citep{hwang00,willingale02}.  The western region suffers from a
higher interstellar column density that attenuates the low energy
X-ray emission, so that the Fe L emission (energies near 1 keV) is
prominent only in the southeast and northwest.  The southeast region
is of particular interest in that the Fe emission here is located
exterior to that of Si.

The left panel of Figure 1 shows the southeast region of Cas A.  The
blast-shocked material may be seen in places as faint, thin arcs at
the outer boundary of the remnant, while the ejecta make up the bright
irregular ring of emission.  The faint, linear filaments of the ejecta
``jet'', which are also seen optically, can be seen towards the top of
the image.  East of the bright ejecta ridge running southward from the
jet, the fainter ejecta emission is largely Fe-rich.  The Fe features
have various morphologies ranging from compact knots to elongated
knots, to relatively diffuse features.

Regions of the remnant that have strong line emission relative to the
underlying continuum may be selected independently of surface
brightness by forming an estimate of the line-to-continuum ratio at
each position.  For the Fe L blends, it is difficult to reliably
subtract the true continuum, but in order to identify the Fe-richest
regions, we have used spectral regions on either side of the L blend
to make a rough estimate of the underlying continuum.  The right panel
of Figure 1 shows the Chandra image in the Fe L blends overlaid with
smoothed contours showing the regions with prominent Fe L
line-to-continuum ratios.  It is seen that the strongest Fe line
emission comes from a line of compact knots running due east at DEC
$\sim 58^\circ 46' 40''$ as well as from diffuse regions of relatively
low surface brightness about an arcminute to the south.  These are
thus our primary target regions to search for sites of $\alpha$-rich
freeze out.  The spectral extraction regions (generally a few
arcseconds in extent) are shown in the left panel of Figure 1.

The level of surface brightness contrast seen for the Fe and Si
features is illustrated in Figure 2.  The first panel shows a vertical
cut through two representative Fe knots (numbered 13 and 14, starting
with 10 closest to the center\footnote{The radial series of knots is
the extension eastward of the east series of knots studied in
\citet{lamhwa}; thus the numbering scheme here starts where that
series left off at A10, increasing eastward to A17.}).  The knots are
seen to sit on a shelf of emission that rises above the general
background.  The second panel shows a vertical cut through the larger
diffuse Fe-rich cloud.  It too sits on a shelf of even fainter
emission, but the surface brightness contrasts are seen to be
significantly lower than for the other Fe-rich knots.  In comparison,
typical Si-rich emission features are much brighter than either the
background or the Fe-rich emission, as is seen in the remaining two
panels of Figure 2.

\subsection{X-ray Spectral Analysis}

We carried out spectral fits for the regions indicated in Figure 1
with a model characterized by a single temperature and single
ionization age, wherein the lightest element included is O, and O
provides the bulk of the continuum.  This is consistent with the
expectation that O is the primary constituent of the ejecta in Cas A,
and follows the approach first adopted by \citet{vink96} for modelling
the X-ray ejecta spectra with an O-rich plasma.  The most prominent
features in the spectra we consider are, however, from Fe.  We also
consider models where Si, rather than O, is the lightest element.  The
background spectrum is generally taken from positions well off the
source.  For the Si/Fe ejecta models, we also take a more localized
background from low-surface brightness regions in the eastern part of
the remnant near the Fe knots.  The results of all these fits are
summarized in Table 1, and the fits for a typical knot are shown in
Figure 3.

For the Fe knots (A series), the fits using the standard off-source
background are generally better with O ejecta than with Si ejecta.  In
both cases, the fitted ionization ages are $\sim 10^{11}$ cm$^{-3}$s
or higher, while the O continuum fits give temperatures that are
higher by a few tens of percent.  The use of a local background for
the Si/Fe continuum fits naturally lowers the fit statistic per degree
of freedom because the local background is less precisely determined
than the standard background, but the actual fitted parameters are not
much changed.

The spectrum of the local background and of the regions surrounding
the compact Fe-rich knots do show enhanced emission from Fe, with
fitted Fe/Si abundance ratios of roughly 1$-$1.5 times solar.  These
Fe/Si ratios are significantly higher than in the Si-rich knots
studied by \citet{lamhwa}---which are nearly devoid of Fe---but not as
high as in the compact Fe knots in Table 1; the fitted Fe/Si
ratio increases for regions closer to the compact knots.  The fitted
ionization ages are several times $10^{10}$ to $10^{11}$
cm$^{-3}$ s, which is within a factor of 2-3 of most of the ionization
ages given in Table 1 for the compact knots.


The faint c6a region lying southward of the knots in the A series is
the most interesting of the Fe-rich features.  For this faint feature
we show the fits using the local background only, as the off-source
background gives statistically unacceptable fits.  The spectrum is
well-fitted with a model including only the elements Si and Fe (plus
Ni), provided that the two elements can have different values of the
ionization age.  The temperature and ionization age for Fe are similar
to those obtained for the Fe-rich knots in Table 1, while the Si
ionization age is similar to that obtained in \citet{lamhwa} for the
O/Si knots.  We also performed a simultaneous fit of epoch 2000 and
2002 ACIS observations for this feature, allowing the two spectra to
have different amounts of detector absorption according to their
observation date.  The parameters obtained are all very close to those
obtained with the epoch 2000 observation alone, although with slightly
smaller error ranges.  Both sets of fits for the c6a region are shown
in Figure 4, and the results summarized in Table 2.  The Si and Fe
ionization ages are not consistent with each other in either of the
fits; their equality can be excluded at higher than 99\% confidence
from the error contours shown in Figure 4 for the joint fit.

The line-to-continuum ratio images in Figure 1 show that the regions
near c6a having very slight surface brightness enhancements over the
background should have high Fe line strengths.  We therefore extracted
source and background spectra for this entire region by specifying
appropriate surface brightness cuts on the slightly smoothed image.
The resulting background-subtracted spectrum can be fitted with a
single component model including only Si and Fe (and Ni), but does not
necessarily require Si and Fe to have separate ionization ages.  An
enhancement of the Fe/Si abundance ratio at a level comparable to that
found for the Fe knots is indicated.  Thus, while this entire region
is significantly Fe-enriched, on the whole it is not as pristine in Fe
as is c6a.  We have examined a number of other features in this
region, and while many of them show Fe enrichments of the order of the
Fe knots, we have so far identified no others where the Fe is a pure
as in c6a.

The Fe-rich knots considered here have Fe/Si abundance ratios that are
higher, generally by an order of magnitude or more, than the typical
0.25 ratio that is obtained through incomplete explosive Si-burning in
a 20 M$_\odot$ progenitor \citep[see Table 2 in][]{thielemann96}.  It
is clearly indicated that some of the Fe is produced with complete Si
exhaustion, as is already well-established, both by earlier X-ray
spectral observations \citep{hughes00} and by the confirmed detection
of the decay products of $^{44}$Ti, which is formed by $\alpha$-rich
freeze out (complete Si-burning that occurs at lower densities).  The
explicit identification of a possible pure cloud of Fe ejecta in
region c6a suggests strongly that this may be one of the sites where
$\alpha$-rich freeze out occurred in Cas A.

The fitted ionization ages for the regions surrounding the A series of
knots in the east are lower than, but of the same order of magnitude
as that in the knots themselves.  A density contrast of a similar
amount (a factor of 2-3) may still exist without triggering the
hydrodynamic instabilities in a manner that destroys the
knots. \citet{lamhwa} demonstrate that the curves of $T_e$ against
$n_et$ are essentially unchanged for such small degrees of clumping.
The Fe/Si abundances of $\sim$solar for this region are also at least
a factor of two lower than the average Fe/Si abundance in the compact
knots.  It seems reasonable to infer that the Fe in the gas
surrounding the compact knots was stripped from Fe-rich knots that
initially had higher density contrasts compared to their surroundings
and therefore did not survive the action of instabilities. Another
possible interpretation would be that the abundances surrounding the
knots are the result of incomplete explosive Si burning, compared with
complete explosive Si burning in the compact knots
themselves \citep[see e.g.][]{thielemann96}.

\section{Fe Ejecta Models and Discussion}

In connection with Fe ejecta knots, it is also necessary to consider
that $^{56}$Fe is originally formed as $^{56}$Ni, which decays
radioactively.  If the initially formed Ni clumps are sufficiently
large, they will be opaque to the $\gamma$ ray radiation produced by
the decays, and thereby expand to form a hot, low-density Fe bubble.
As the reverse shock traverses these bubbles, the resulting turbulence
plays an important role in mixing the Fe ejecta into overlying ejecta
layers \citep{li93}, as well as causing filamentation in the overlying
(Si-rich) ejecta \citep{blondin01}.  The Fe associated with the bubble
effect should be characterized by diffuse morphologies and low
ionization ages.  The relative compactness of an Fe ejecta knot seen
in the Type Ia (thermonuclear runaway) remnant of Tycho's supernova,
caused \citet{wang01} to suggest that the Fe in question there is not
$^{56}$Fe, but rather $^{54}$Fe, which is not formed by radioactive
decays and is therefore not subject to the bubble effect.  For
core-collapse supernovae, and for Cas A in particular, this is a less
satisfactory explanation because it is then difficult to produce a
sufficient mass of $^{54}$Fe, particularly if $\alpha$-rich freeze out
occurred during the explosion.  The bubble mechanism has also been
suggested to be connected with the ring-like filamentation seen in the
optically emitting ejecta (Fesen 2001), and indeed may be
responsible for the highly filamentary X-ray emission from Si as well.
Presumably any Fe associated with the bubble effect is now too faint
and underionized to be readily identifiable. On the other hand, if
some of the $^{56}$Ni clumps were sufficiently small at the outset,
these could become optically thin to the $\gamma$-ray radiation
sufficiently early on that the bubble effect would be minimized for
these clumps.  These are the Fe clumps that we see today.

We can demonstrate that these Fe clumps were indeed small enough to
minimize the bubble effect.  The observed clump diameters of about
3$''$ give present day radii of $7.5\times 10^{16}$ cm at the 3.4 kpc
distance of Cas A.  Assuming that the radius increased proportionally
to time, we get $r=6\times 10^{11}\ t_{\rm days}$ cm for the radius of
a $^{56}$Ni/$^{56}$Co clump at time $t_{\rm days}$ days after the
explosion.  The clump electron density varies as $t^{-3}$ until
interaction with the reverse shock, so extrapolating our $n=7$ model
back in time gives an electron density in the range
$n_e=\left(1.5-3\right)\times 10^{14}/t_{\rm days}^3$ cm$^{-3}$.  The
optical depth to the center of the clump is then \citep{li93}
\begin{equation}
\tau\left(t_{\rm days}\right) \simeq \sigma n_er\simeq \left(18-36\right)/t_{\rm days}^2,
\end{equation}
where $\sigma = 0.31\sigma _{\rm T}$ for $\gamma $ rays of energy
about 1 MeV, and $\sigma _{\rm T}$ is the Thomson cross section. Hence
we estimate that these clumps became optically thin ($\tau\sim1$) to
their own $\gamma $ rays about 5 days after explosion. This is
significantly shorter than the $^{56}$Ni and $^{56}$Co lifetimes of
8.8 and 111.3 days, respectively, so most of the radioactive energy is
deposited outside the clump.  We therefore do not expect that the Fe
knots we observe today have undergone any type of bubble expansion.

We pursue a more quantitative interpretation of the Fe knots in Cas A
than has been previously attempted, by calculating the variation of
the electron temperature $T_e$ with ionization age $n_et$ for a
variety of ejecta density profiles, as described in \citet{lamhwa}. We
concentrate on models with an explosion energy of $2\times 10^{51}$
ergs, shocked ejecta mass $2 M_{\sun}$, and circumstellar density
$\rho r_b^2=14$ pc$^2$ cm$^{-3}$, with $r_b$ the radius of the blast
wave in pc. \citet{lamhwa} treated the case of pure O ejecta only;
here we take compositions by mass of O:Si:Fe = 0.83:0.06:0.11 and
Fe:Si = 0.9:0.1, which are approximately consistent with the spectral
fits to the observed knots assuming an O-rich and Fe dominated
composition for the continuum, respectively.  These models are shown
in Figures 5 and 6 as solid lines; the fit results are superposed as
crosses for the ``A'' series of knots and, in Figure 6 only, as boxes
for the diffuse Fe clouds, with the symbol size indicating the fit
uncertainties in $T_e$ and $n_et$. Table \ref{tbl-3} gives the ejecta
mass coordinates inferred for each knot by matching the fitted value
of $n_et$ to the corresponding location in the ejecta for the
hydrodynamic model. In this, we assume that the clumps have the same
density as their surroundings, and infer electron densities of 10-50
cm$^{-3}$ for the clumps. If the knots are actually overdense, the
$n_et$ we fit would correspond to a later reverse shock passage and
hence place the knots further in in the ejecta.  We estimate that a
factor of three overdensity would lead to an overestimate of $q$ by
around 0.06; a similar underdensity would place the knots further out
in the ejecta, by a similar amount. Following reverse shock passage,
our Lagrangian plasma parcel likely undergoes interactions with
secondary shocks while the reverse shock is relatively nearby, leading
in any case to uncertainties in $q$ of a similar magnitude.

The mass of $^{56}$Fe contributed by the various knots is computed
from the fitted emission measure and element abundances of the knots
in Table \ref{tbl-3}.  Spherical geometries were assumed for the
knots, and boxes for the diffuse features.  The corresponding electron
densities are also given in the table, and for comparison, the
electron densities that are inferred from the models.  The agreement
between the observationally and theoretically inferred densities is
generally within a factor of two, though it is somewhat worse for the
larger East region.  In any case, the observationally determined
electron densities are systematically lower than the theoretically
determined ones, which would seem to reinforce our assumption that the
knots are not overdense.

According to the hydrodynamical model in \citet{lamhwa}, the reverse
shock is currently at an ejecta mass coordinate $q=0.1-0.14$ for
$n=9-7$ models respectively. Hence the inner 10\% of ejecta, i.e. the
inner 0.2 $M_{\sun}$, has yet to encounter the reverse shock. Cas A is
highly unlikely to have ejected more than 0.05 - 0.1 $M_{\sun}$ of
$^{56}$Ni (which $\beta$ decays to $^{56}$Fe), so the reverse-shocked
$^{56}$Fe that we do see must have been mixed out into the envelope by
Rayleigh-Taylor instabilities shortly after explosion.  The estimated
knot masses in Table \ref{tbl-3} probably amount to a few percent of
the total mass ejected, which is similar to the mass of $^{56}$Ni
inferred to have been mixed out into the envelope of SN 1987A
\citep{pinto88}.

According to the mass coordinates we infer, the Fe clumps exist out to
about 0.4 of the 2 $M_{\sun}$ ejecta, or 0.8 $M_{\sun}$. Adding 1.3
$M_{\sun}$ for the mass of the compact object, we arrive at $0.8 + 1.3
= 2.1\ M_{\sun}$ for the observed outer extent of Ni mixing.  This
places the knots at a mass coordinate corresponding to the (Ni+Si)/O
interface in a star that is initially of about 20 $M_{\sun}$
\citep{woosley95}.  Though this is an appealing inference, it is
necessary to account for possible selection effects that determine the
visibility of the knots.  In the case of an O-rich composition for the
knots, this apparent outer extent appears to be a real outer extent
(see Figure 5).  We expect that if higher $n_et$ knots exist now, we
would be able to see them.  For Fe-dominated composition, however,
knots that encounter the reverse shock earlier than about 35 years
after explosion (at ejecta mass coordinates $q_{\rm FeSi}\gtrsim
0.45$) will have by now cooled by radiative losses to temperatures
below detectability as X-ray knots.  In this case, the observed outer
extent could be limited by this selection effect.  We also fail to
detect Fe knots with $n_et \lesssim 10^{11}$ cm$^{-3}$s, i.e. in the
inner 0.3 of the ejecta, or interior to 1.9 $M_{\sun}$, if the compact
object is included. This might also be a selection effect caused by
the rapid decrease in density as one move inwards into the ejecta.  Fe
that is deep in the ejecta may also be more likely to have created
bubbles due to the radioactivity of its parent $^{56}$Co nucleus, and
so become even less dense than the O-rich ejecta that is assumed to
surround it. The inferred mixing of the ``c6a'' and ``east'' regions
out into regions that were originally O and are burnt to Si during the
explosion, is also consistent with the inference that $^{44}$Ca found
in SiC X grains is a decay product of $^{44}$Ti formed in core
collapse supernovae \citep[e.g.][]{clayton02}.

It appears that the O:Si:Fe composition we infer can only be achieved
by mixing products of complete Si burning with the products of O
burning (to get Fe:Si right), and it is unlikely that a ``knot'' would
survive such mixing as a distinct object. More plausible is the pure
Fe:Si composition, that requires no further mixing after Si
burning. Consequently the selection effect due to the radiative
cooling of Fe rich knots discussed above is probably operative, and
more Fe may exist at lower temperatures further out in the ejecta of
Cas A.

\section{Conclusions}

Using Chandra observations combined with new modelling techniques, we
have constrained the ejecta density profiles and Lagrangian mass
coordinates of some of the Fe-rich knots in the southeastern region of
Cas A. This makes it possible for the first time to compare
observations in a quantitative manner with explosion models, and
allows tests of nucleosynthesis and Rayleigh-Taylor instability in
core collapse supernova explosions. The inference is that Fe and Si in
knots are relatively uncontaminated by lower $Z$ elements, and that
their inferred ejecta mass coordinates appear to be entirely
consistent with the expected Rayleigh-Taylor turbulence in a Type Ib
supernova. We have further identified a region of nearly pure Fe
ejecta that is a promising candidate for a site of $\alpha$-rich
freeze out. Again it is mixed out by Rayleigh-Taylor turbulence, but
as a site $\alpha$-rich freeze out associated with $^{44}$Ti
production, its origin should have been closer to the center and the
mass cut than those of the other Fe rich knots.

It is clearly desirable to make a complete census of the X-ray
emitting Fe ejecta regions in Cas A and identify more regions that are
highly enriched in Fe. For example, the ejected $^{44}$Ti mass is
estimated at $1.8\times 10^{-4} M_{\sun}$ (Vink et al.\, 2001; Vink \&
Laming 2003), and 500-1000 times more $^{56}$Ni by number is predicted
globally from $\alpha$-rich freeze out (Woosley \& Hoffman 1991;
Arnett 1996; Thielemann, Nomoto \& Hashimoto 1996; The et al.\, 1998),
so considerably more $\alpha$-rich freeze out ashes should be present
in Cas A to account for the $^{44}$Ti emission than are inferred to be
in regions ``east'' and ``c6a''. In this work we have principally been
limited by the statistical quality of the current Chandra data sets
when extracting spectra from the smallest possible spatial regions. A
considerably deeper observation is required to take full advantage of
the unprecedented spatial resolution available with Chandra, which
appears to be crucial in studying the Fe emission and making progress
on important issues in supernova physics such as asymmetries,
nucleosynthesis and the location of the mass cut.

\acknowledgements  We wish to thank Larry Rudnick and Tracey Delaney
for communication of their results prior to publication, and for
allowing us access to their second epoch Chandra/ACIS data of Cas A
prior to becoming public.  JML was supported by basic research funds
of the Office of Naval Research.

\begin{deluxetable}{ccccccccc}
\tabletypesize{\scriptsize}
\tablecaption{Fits to Fe-rich Knots\label{tbl-1}}
\tablewidth{0pt}
\tablehead{
\colhead{Knot:\tablenotemark{a}} & \colhead{A10} & \colhead{A11} & \colhead{A12} & \colhead{A13} & \colhead{A14} &
\colhead{A15} & \colhead{A16} & \colhead{A17} \\
  }
\startdata
\cutinhead{Abundances relative to O, standard background}
$k_{\rm B}T_e$\tablenotemark{b}&2.11 & 2.13 & 1.57 & 2.03 & 1.79 & 1.58 & 1.98 & 1.99 \\
(keV) &1.80-2.52 &1.76-2.59 & 1.50-1.67 & 1.40-2.45 & 1.58-2.07 & 1.46-1.87 & 1.71-2.26 & 1.86-2.25 \\
$n_et$                         &1.98 & 2.32 &7.92 & 0.92 & 2.28 & 2.19 & 0.97 & 1.63 \\
($10^{11}$ cm$^{-3}$s) &1.48-3.07 & 1.45-3.21 & 4.98-11.7 & 8.63e+10-1.90 & 1.65-3.72 & 1.74-2.90 & 7.57e+10-1.15 & 1.58-2.05 \\
Mg/O   &0.007 & 0.075 & 0 & 0.085 & 0 & 0.043 & 0.018 & 0.013 \\
 & $<$ 0.20 & $<$0.13 & $<$0.20 & 0.056-0.15 & $<$0.075 & 0.023-0.095 & $<$0.049 & $<$ 0.049 \\
Si/O\tablenotemark{c}            &0.40 & 0.29 & 1.34 &  0.38 & 0.32 & 0.13 & 0.14 & 0.076 \\
 &0.29-0.85 &0.23-0.57 & 0.59-2.48 & 0.26-0.48 & 0.22-0.49 & 0.094-0.19 & 0.12-0.18 & 0.066-0.084 \\
S/O    & 0.43 & 0.30 & 0.97 & 0.34 & 0.28 & 0.17 & 0.14 & 0.072 \\
 &0.30-0.91 & 0.22-0.46 & 0.44-3.20 & 0.26-0.39 & 0.18-0.40 & 0.12-0.20 & 0.11-0.17 & 0.051-0.093 \\
Ca/O   & 0.64 & 0.53 & 2.41 & 0.49 & 0.70 & 0.50 & 0.48 & 0 \\
 &0.21-1.07 & 0.27-0.68 & 1.50-3.11 & 0.11-0.85 & 0.28-1.13 & 0.18-0.83 & 0.16-0.79 & $<$ 0.074 \\
Fe/O                             &1.29 & 0.72 & 2.67 & 0.28 & 1.09 & 0.25 & 0.25 & 0.26 \\
 &0.84-3.20 &0.50-1.83 & 1.30-20.4 & 0.14-0.35 & 0.74-1.98 & 0.17-0.44 & 0.20-0.34 & 0.22-0.29 \\
$N_{\rm H}$                    &1.33 & 1.27 & 1.37 & 1.16 & 1.42 & 1.17 & 1.20 & 1.18 \\
($10^{22}$ cm$^{-2}$) &1.19-1.39 &1.17-1.40 &1.33-1.40 &1.05-1.20 & 1.35-1.48 & 1.03-1.38 & 1.12-1.29 & 1.05-1.26 \\
EM ($10^{11}$ cm$^{-5}$)       & 2.25 & 4.68 & 1.69 & 3.23 & 2.88 & 4.39 & 4.26 & 6.86 \\
$\chi ^2$                      &116.4 & 218.2 & 180.0 & 157.3 & 129.9 & 78.2 & 135.1 & 146.7 \\
$\chi ^2$/dof                  &1.25 & 1.93 & 1.70 & 1.69 & 1.44 & 0.98 & 1.54 & 1.47 \\
\# dof                         & 93  & 113  & 106  & 93   & 90   & 80 & 88  & 100 \\
\# cts                         & 5221 & 6748  & 5541  & 4679  & 5123  & 3702 & 4396 & 5416 \\
\cutinhead{Abundances relative to Si, standard background}
$k_{\rm B}T_e$\tablenotemark{b}& 1.73 & 1.59 & 1.56 & 1.23 & 1.50 & 1.32 & 1.20 & 1.47 \\
(keV) & 1.65-1.84 & & 1.48-1.61 & & 1.42-1.57 & 1.32-1.40 & & \\
$n_et$                 & 3.69 & 7.55 & 10.1 & 4.56 & 5.59 & 10.9 & 9.57 & 7.61 \\
($10^{11}$ cm$^{-3}$s) & 3.41-5.48 & & $>$7.73 & & 4.12-8.27 & 5.89-16.4 & & \\
Fe/Si                     & 3.55 & 2.84 & 1.91 & 1.43 & 3.67 & 2.25 & 2.10 & 3.85 \\
 & 3.05-3.74 & & 1.54-2.19 & & 3.30-4.10 & 1.95-2.83 & & \\
$N_{\rm H}$            & 1.44 & 1.53 & 1.41 & 1.66 & 1.53 & 1.49 & 1.50 & 1.48 \\
($10^{22}$ cm$^{-2}$)  & 1.35-1.49 & & 1.36-1.49 & 1.50-1.60 & 1.33-1.60 & & \\
$\chi ^2$              & 124.0 & 236.3 & 182.2 & 227.2 & 145.5 & 134.8 & 236.6 & 243.4 \\
$\chi ^2$/dof          & 1.32 & 2.07 & 1.70 & 2.42 & 1.60 &1.66 & 2.66 & 2.41 \\
\# dof                 & 94 & 114 & 107 & 94 & 91 & 81 & 89 & 101 \\
\cutinhead{Abundances relative to Si, local background}
$k_{\rm B}T_e$\tablenotemark{b}& 1.77 & 1.68 & 1.58 & 1.24 & 1.52 & 1.34 & 1.20 & 1.47 \\
(keV) & 1.63-1.95 & 1.60-1.78 & 1.52-1.69 & 1.07-1.92 & 1.46-1.62 & 1.29-1.36 & & \\
$n_et$                 & 3.65 & 7.40 & 1.02e+12 & 3.41 & 5.33 & 8.33 & 9.58 & 7.97 \\
($10^{11}$ cm$^{-3}$s) & 2.83-5.26 & 5.30-10.0 & $>$6.85 & 1.60-5.58 & 3.75-6.80 & 6.18-14.7 & & \\
Fe/Si                     & 3.84 & 2.76 & 1.95 & 0.39 & 3.93 & 2.57 & 2.10 & 3.89 \\
 & 3.27-4.38 & 2.32-3.38 & 1.52-2.28 & 0.28-0.49 & 3.50-4.58 & 1.66-2.97 & & \\
$N_{\rm H}$            & 1.44 & 1.45 & 1.42 & 0.93 & 1.54 & 1.53 & 1.50 & 1.48 \\
($10^{22}$ cm$^{-2}$)  & 1.33-1.50 & 1.34-1.52 & 1.38-1.49 & 0.69-1.23 & 1.47-1.60 & 1.35-1.54 & & \\
$\chi ^2$              & 109.4& 218.1& 170.6 & 179.3 & 124.8 & 111.2 & 210.3 & 209.1 \\
$\chi ^2$/dof          & 1.16 & 1.91 & 1.59 & 1.91 & 1.37 & 1.37 & 2.36 & 2.07
\\
\# dof                 & 94   & 114  & 107 & 94 & 91 & 81 & 89 & 101 \\
\enddata

\tablenotetext{a}{Knots are numbered moving out to the limb, continuing the east radial series in Paper I.  Errors quoted are 90\% confidence when the $\chi^2/dof < 2.0$.}
\tablenotetext{b}{Electron temperature in keV. Uncertainties are typically $\pm 0.1$ keV.}
\tablenotetext{c}{Element abundance ratio by number relative to solar values of
\citet{anders89}: O/H = 8.51e-4, Mg/H=3.80e-5, Si=3.55e-5, S=1.62e-5, Ar=3.63e-6, Ca=2.29e-6, Fe=4.68e-5, Ni=1.78e-6. Note that these are superseded by \citet{grevesse98}, \citet{allende01}, and \citet{allende02}. In particular abundances relative to O increase by 1.75.  Where abundances are given relative to a particular element, such as O or Si, the abundance of that element was fixed at its solar value for the fits.}
\end{deluxetable}

\begin{deluxetable}{cccc}
\tabletypesize{\scriptsize}
\tablecaption{Fits to Diffuse Regions\label{tbl-2}}
\tablewidth{0pt}
\tablehead{
\colhead{Region:} & \colhead{c6a} & \colhead{c6a (joint)} & \colhead{East} \\
}
\startdata
$k_{\rm B}T_e$\tablenotemark{a} & 2.06 & 1.90 & 2.59 \\
(keV) & 1.88-2.30 & 1.80-1.97 &  2.09-2.97 \\
$n_et_{Si}$ & 0.74 & 0.80 & 1.59  \\
($10^{11}$ cm$^{-3}$s) & 0.187-1.71 & 0.421-1.30 & 1.26-2.32 \\
S/Si\tablenotemark{b} & 0 & 0 & 0.72 \\
  & ... & ... & 0.55-0.90 \\
Ar/Si & 0 & 0 & 2.15 \\
  & ... & ... & 1.16-3.16 \\
Fe/Si & 22.8 & 20.0 & 5.77 \\
 & 16.1-49.4 & 14.1-27.1 &5.43-6.78  \\
$n_et_{Fe}$ & 3.3 & 3.81 &  ... \\
($10^{11}$ cm$^{-3}$s) & 2.5-4.0 & 3.32-4.90 & ... \\
$N_{\rm H}$ & 1.34 & 1.41 & 1.27 \\
($10^
{22}$ cm$^{-2}$) & 1.27-1.43 & 1.35-1.47 & 1.23-1.31\\
EM (cm$^{-5}$) & 1.54e+10 & 1.89e+10 & 1.60e+11\\
$\chi ^2$ & 110.3 & 262.7 & 244.8 \\
$\chi ^2$/dof & 1.33 & 1.45 & 1.56 \\
\# dof & 83 & 181 & 157 \\
\# cnts\tablenotemark{c}& 5668 (60\%) & 5773+5608 (60\%) & 27816 (57\%) \\
Region Size & & & \\
\enddata
\tablenotetext{a}{Electron temperature in keV. Uncertainties are typically $\pm 0.1$ keV.}
\tablenotetext{b}{Abundance by number relative to Si, which is fixed at the solar value.}
\tablenotetext{c}{Quantity in parentheses is the fraction of the counts remaining after background subtraction.}
\end{deluxetable}

\begin{deluxetable}{cccccc}
\tabletypesize{\scriptsize}
\tablecaption{Ejecta knot abundances and mass coordinates\label{tbl-3}}
\tablewidth{0pt}
\tablehead{
\colhead{Region} &
\colhead{$q_{\rm OSiFe}$\tablenotemark{a}} &
\colhead{$q_{\rm FeSi}$\tablenotemark{b}} &
\colhead{Fe masses\tablenotemark{c}}& \colhead{Obs. $n_e$\tablenotemark{d}}& \colhead{Model $n_e$\tablenotemark{e}}\\}
\startdata
A10 & 0.32-0.37& 0.35-0.39&  7e-5 & 12 & 12-20\\
A11 & 0.32-0.37& 0.39-0.44&  8e-5 & 13 & 20-39\\
A12 & 0.38-0.44& $>0.41$  &  9e-5 & 15 & $>26$\\
A13 & 0.29-0.34& 0.31-0.38&  4e-5 & 7 & 7-18\\
A14 & 0.31-0.36& 0.37-0.41&  8e-5 & 12 & 16-26\\
A15 & 0.31-0.34& 0.40-0.46&  4e-5 &7 & 23-50\\
c6a & - & 0.36-0.38 &  2e-4 &8 & 14-18\\
East & - & 0.31-0.34 &  $\sim$1e-3 & 2 & 7-11\\
\enddata
\tablenotetext{a}{Ejecta mass coordinate for OSiFe composition, assuming no
over/under density.}
\tablenotetext{b}{Ejecta mass coordinate for FeSi composition, assuming no
over/under density.}
\tablenotetext{c}{Fe masses computed from fitted emission measures and Fe abundances assuming pure Fe, and spherical knots of
1.75$''$ radius, except for c6a (3.8$''\times 8.8''$ box) and East (11$'' \times 24''$).}
\tablenotetext{d}{Electron densities in cm$^{-3}$ from emission measure and fitted abundances.}
\tablenotetext{e}{Electron densities in cm$^{-3}$ from $n=7$ model corresponding to the range in $q_{\rm FeSi}$.}
\end{deluxetable}

\clearpage

\begin{figure}
\plottwo{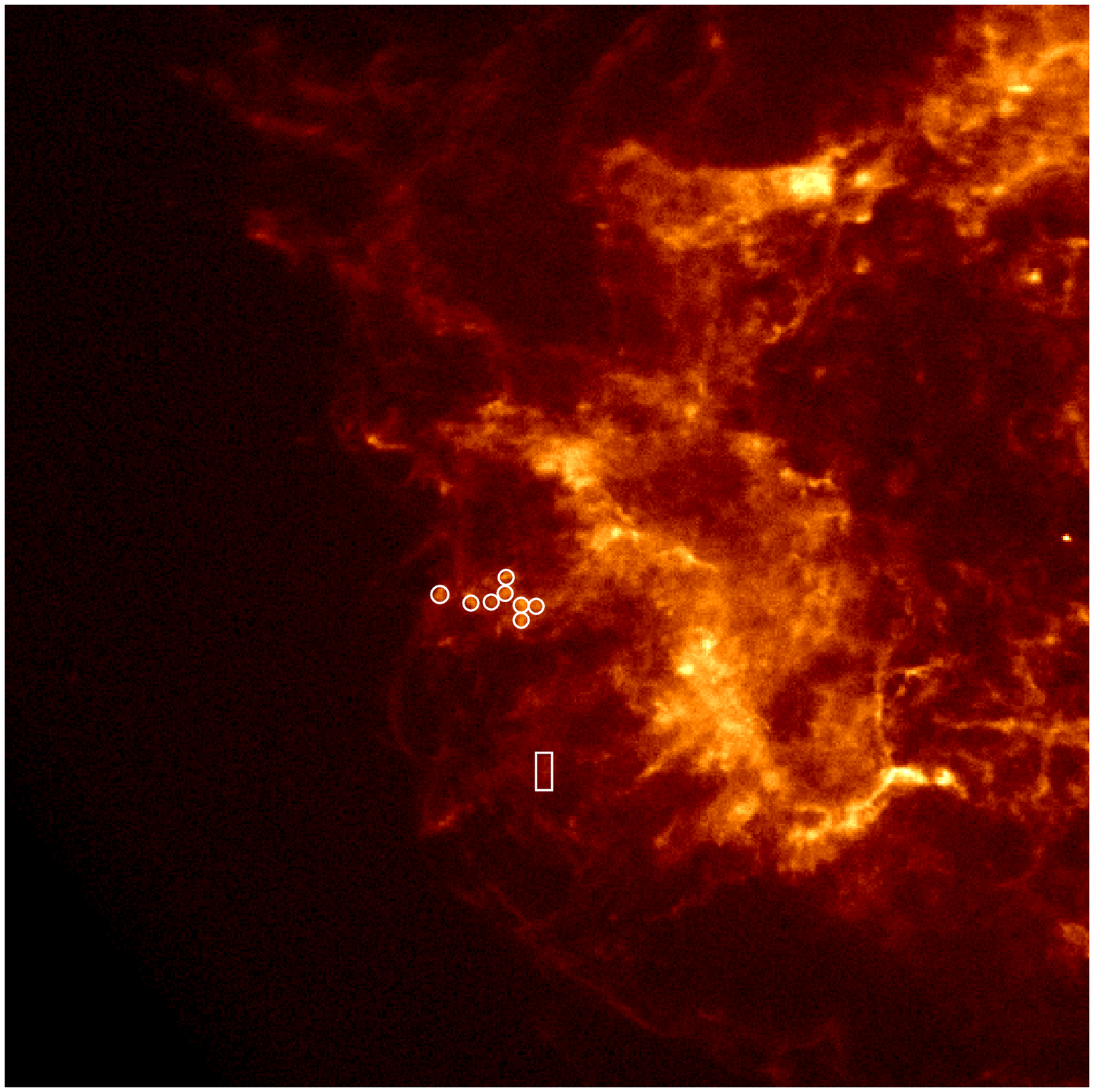}{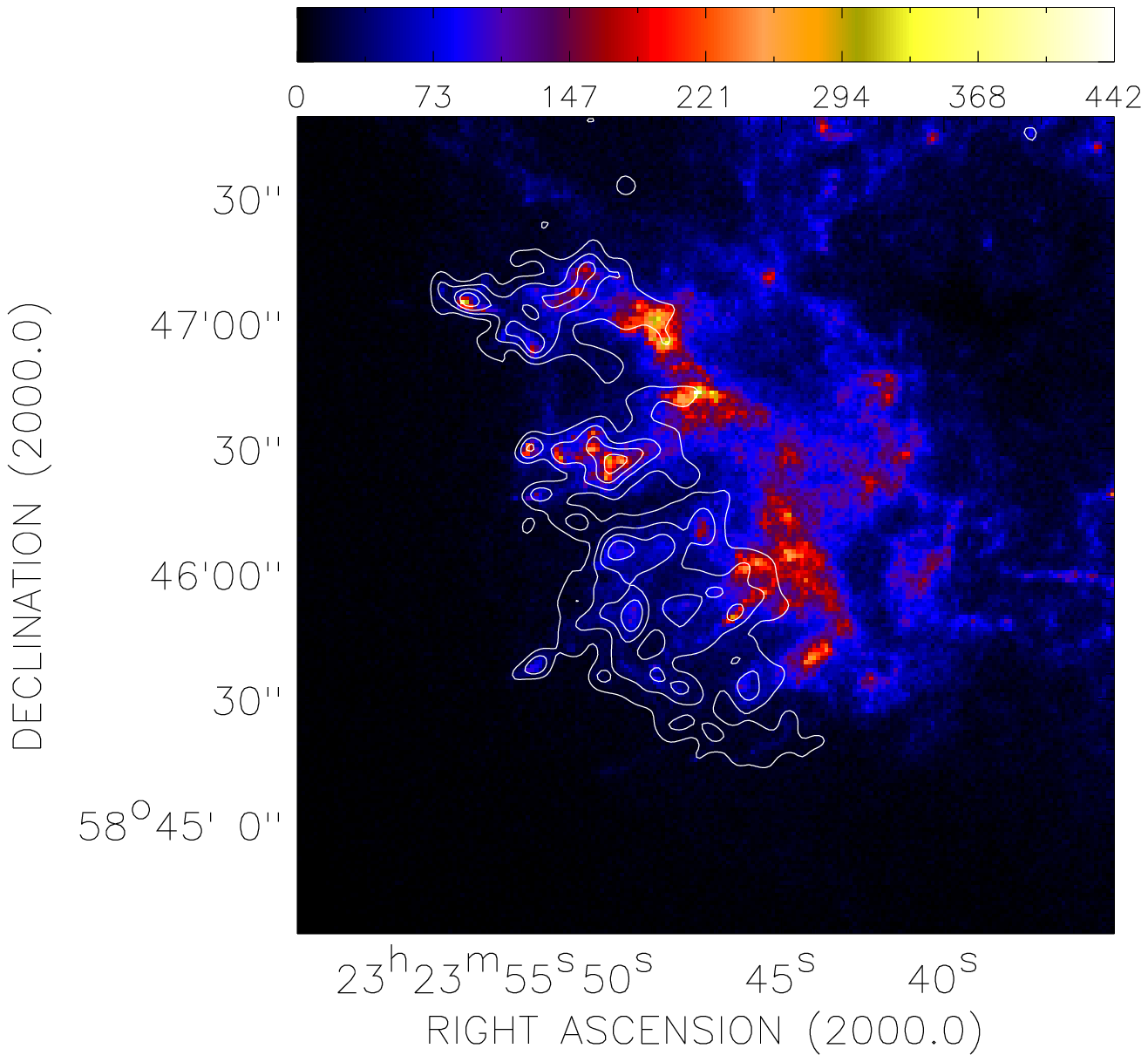} \figcaption{Detail of the southeastern
region of Cas A as imaged by the Chandra X-ray Observatory in January
2000.  The spectral extraction regions used for the linear series of
compact Fe-rich knots (circles, starting from the center moving
outwards) and the extremely Fe-rich cloud (box) are marked.  On the
right is the Chandra image of roughly the same region, but including
only the photons in the Fe L energy range; superposed are contours
showing Fe L line-to-continuum ratios.}
\end{figure}

\begin{figure}
\centerline{\includegraphics[scale=0.50]{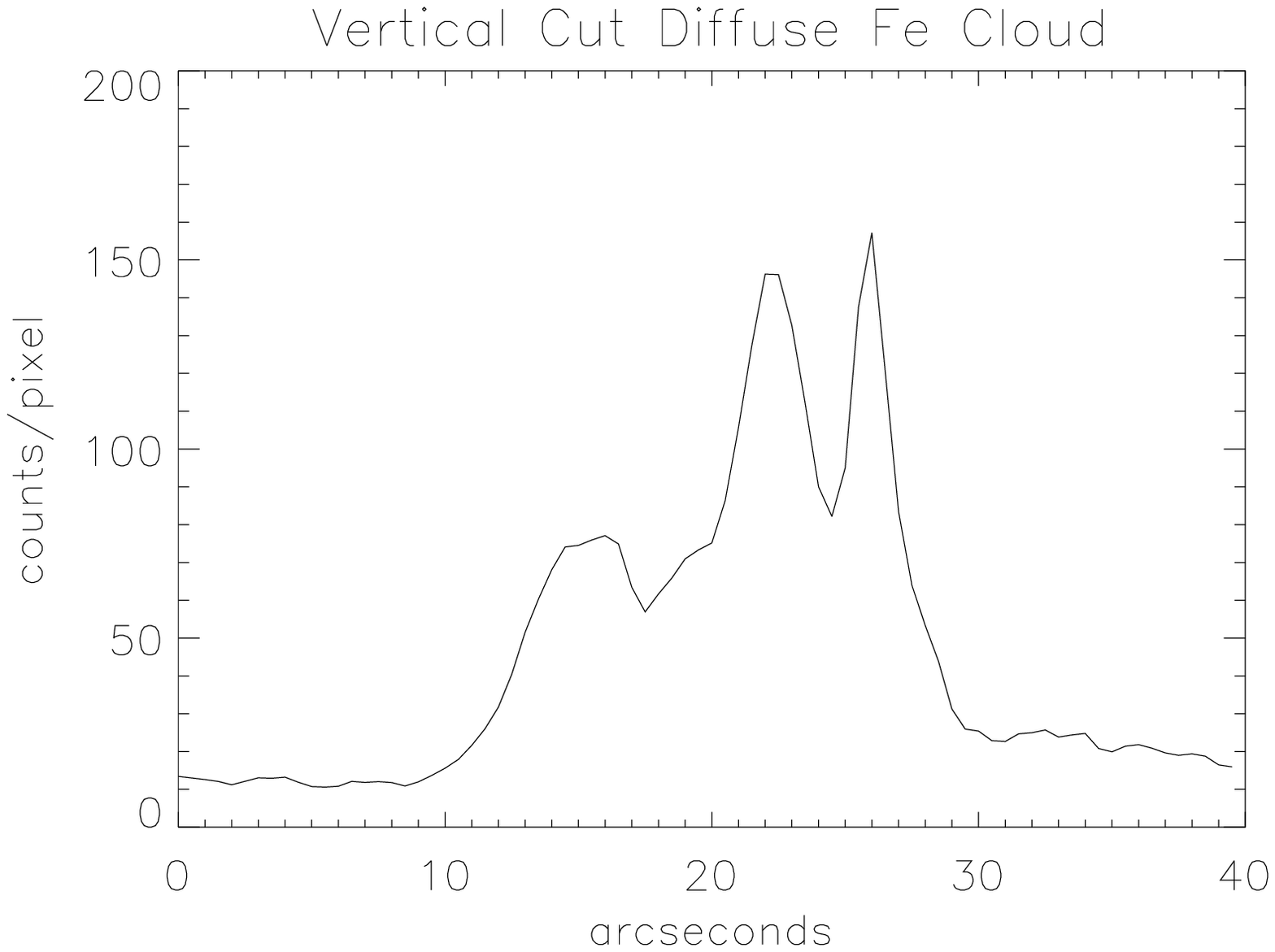}
\includegraphics[scale=0.50]{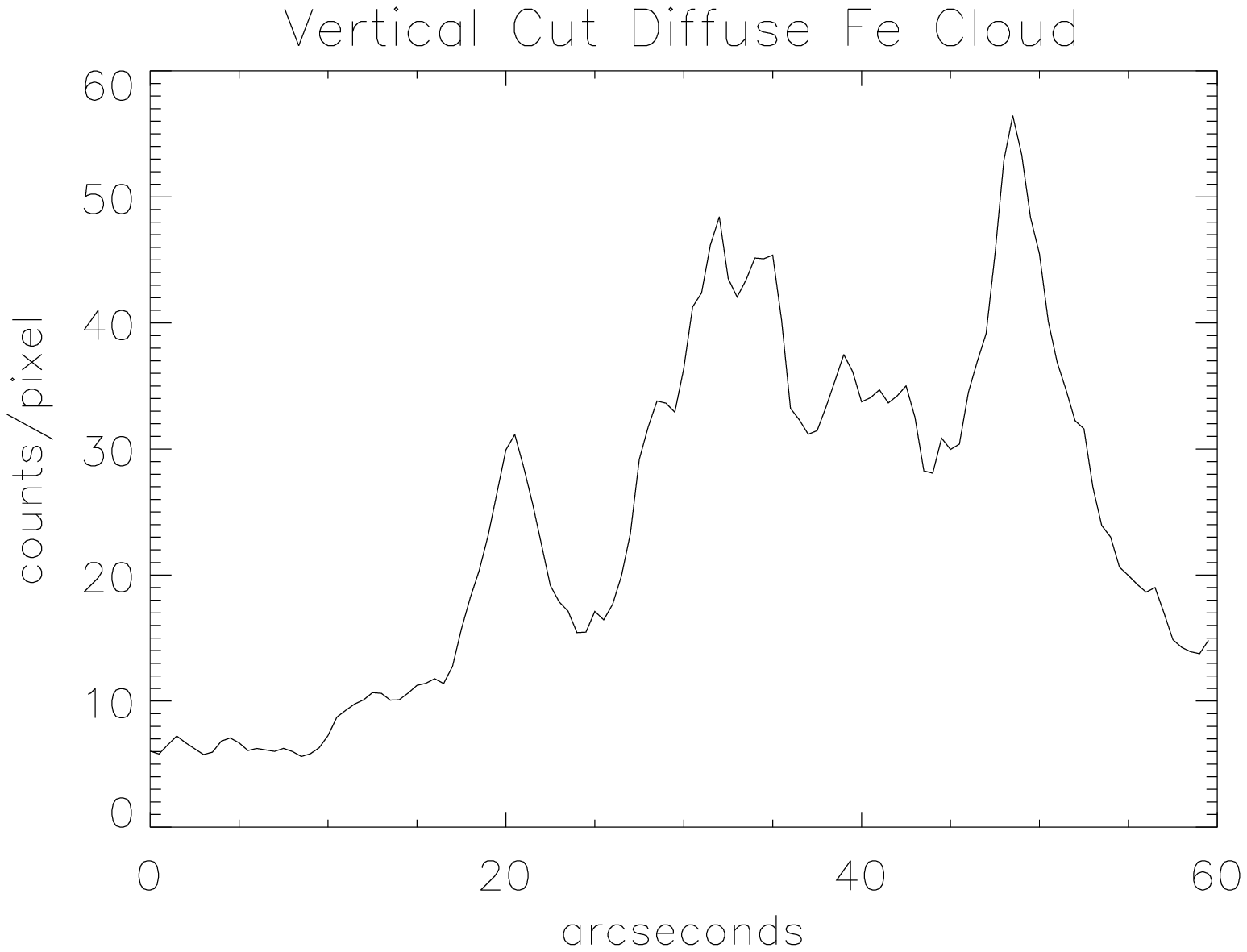}}
\centerline{\includegraphics[scale=0.50]{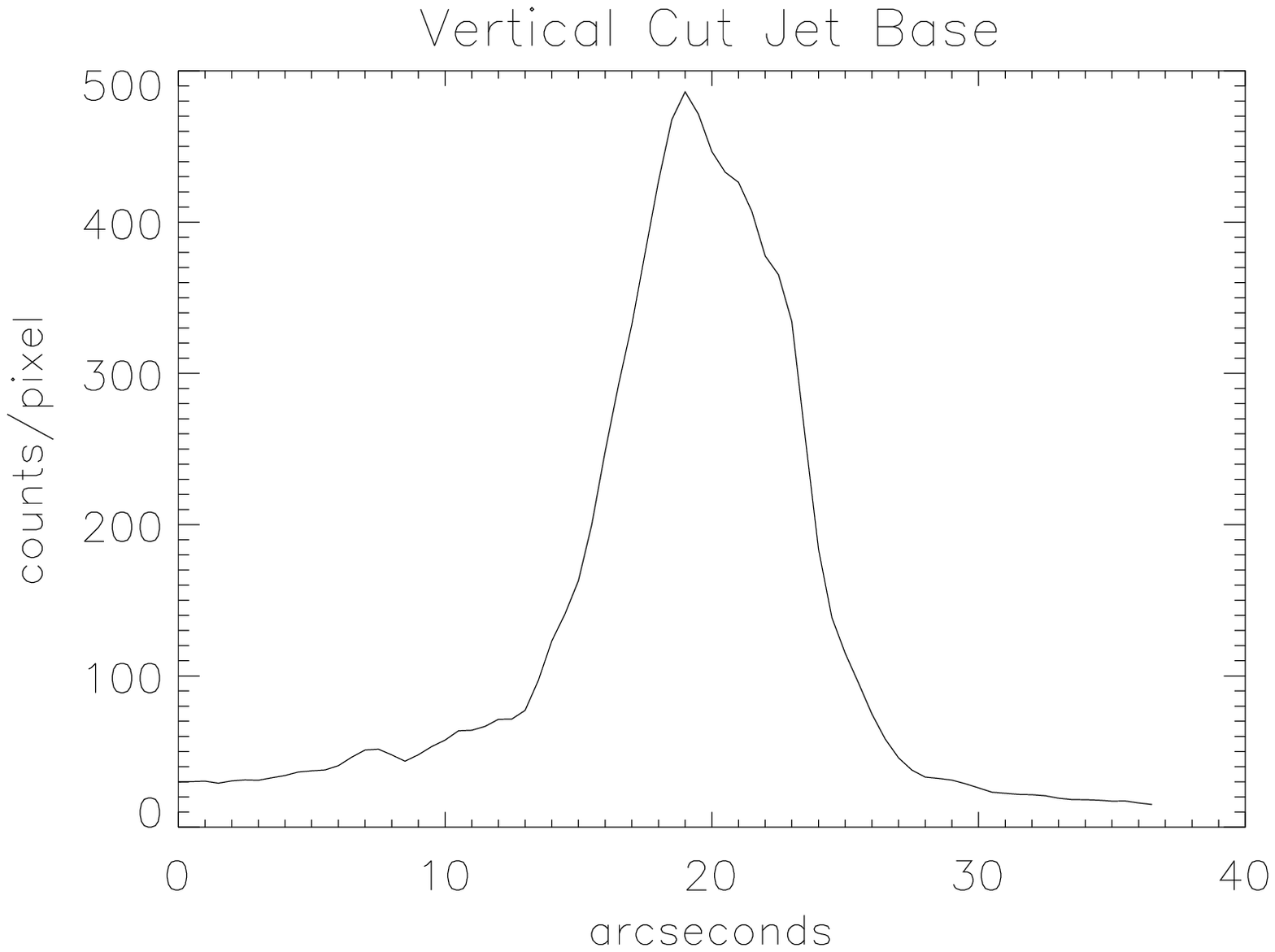}
\includegraphics[scale=0.50]{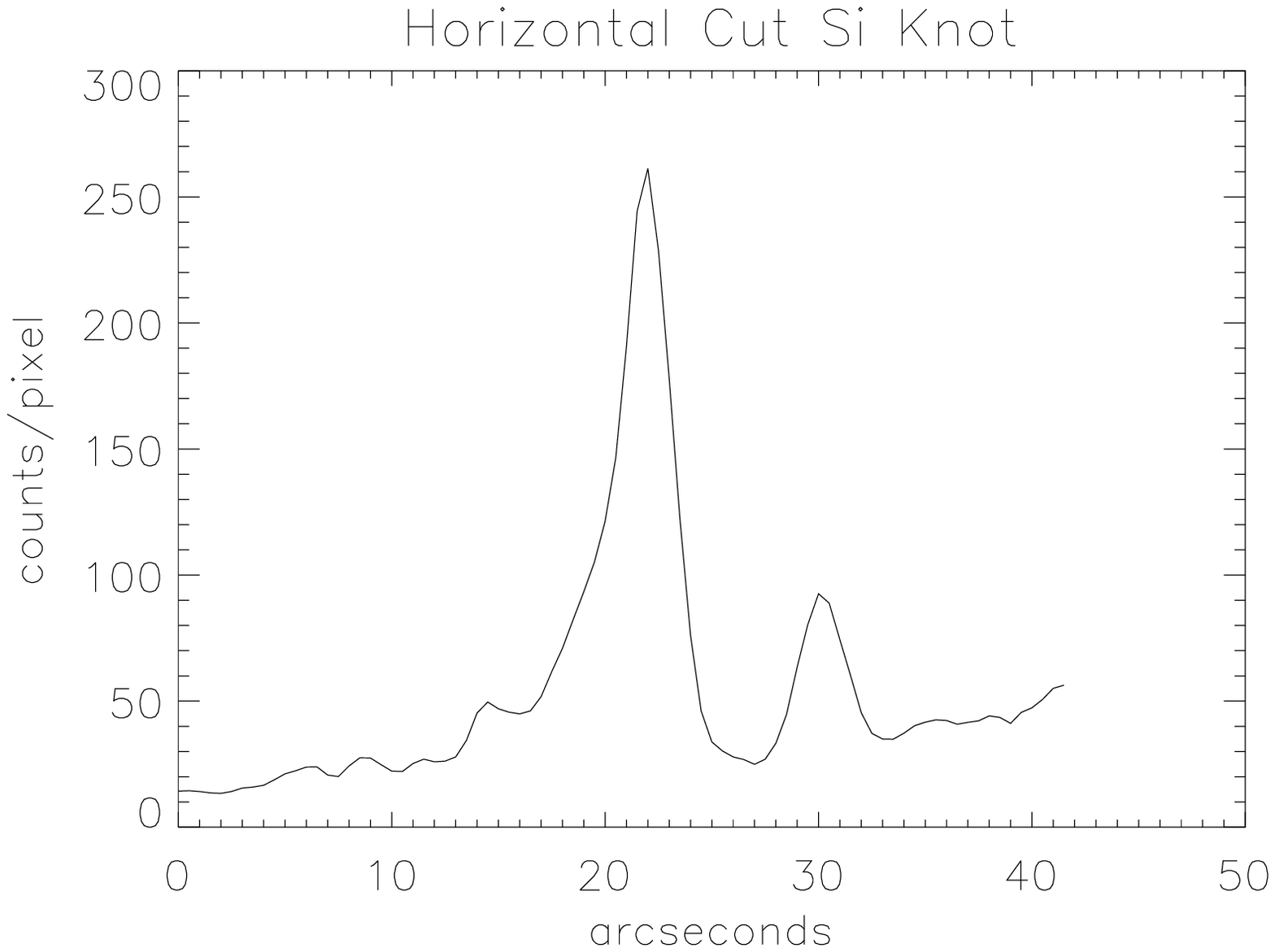}}
\figcaption{Crosswise cuts across various features show the relative
surface brightness contrast.  Plotted are the average counts per pixel
across (a) a 3$''$-wide vertical strip through Fe-rich knots 13 and
14 (b) a 4$''$-wide vertical strip passing through the very Fe-rich
diffuse cloud (c) a 6$''$-wide vertical strip through the base of the
jet, analyzed in Paper I (d) a 3$''$-wide horizontal strip through an
isolated Si-rich knot.}
\end{figure}

\begin{figure}
\centerline{\includegraphics[scale=0.35,angle=-90]{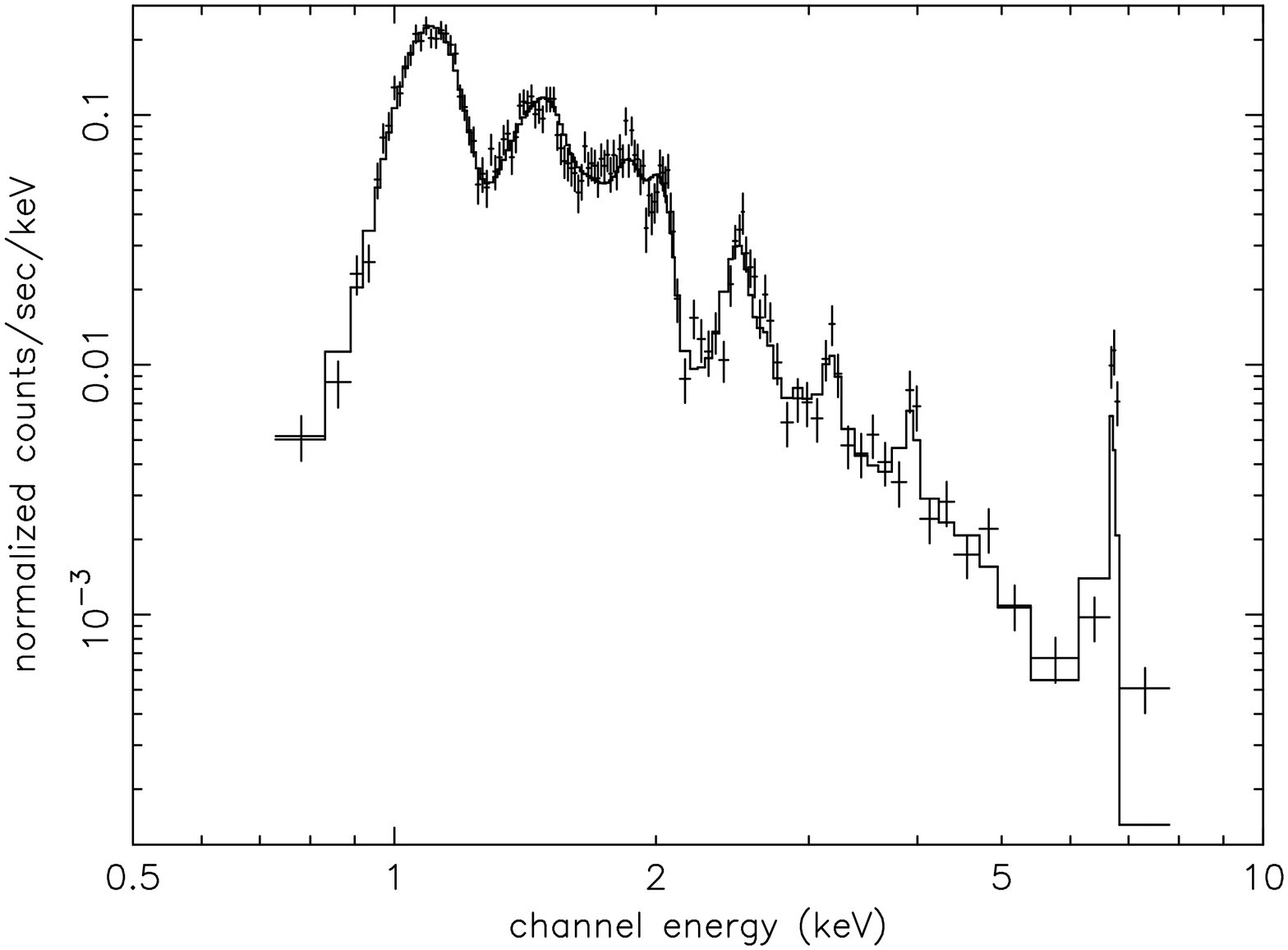}
\includegraphics[scale=0.35,angle=-90]{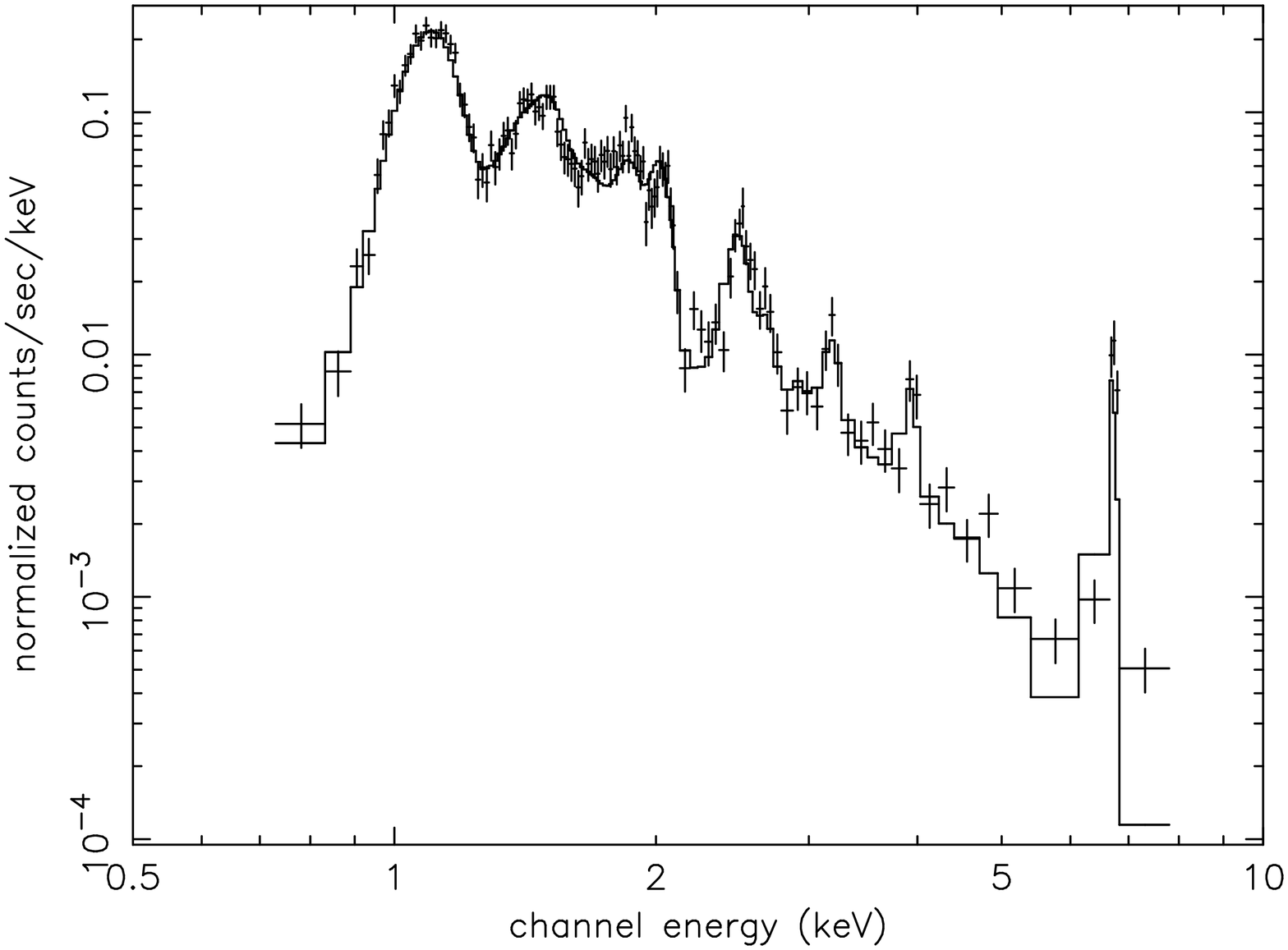}}\bigskip
\centerline{\includegraphics[scale=0.35,angle=-90]{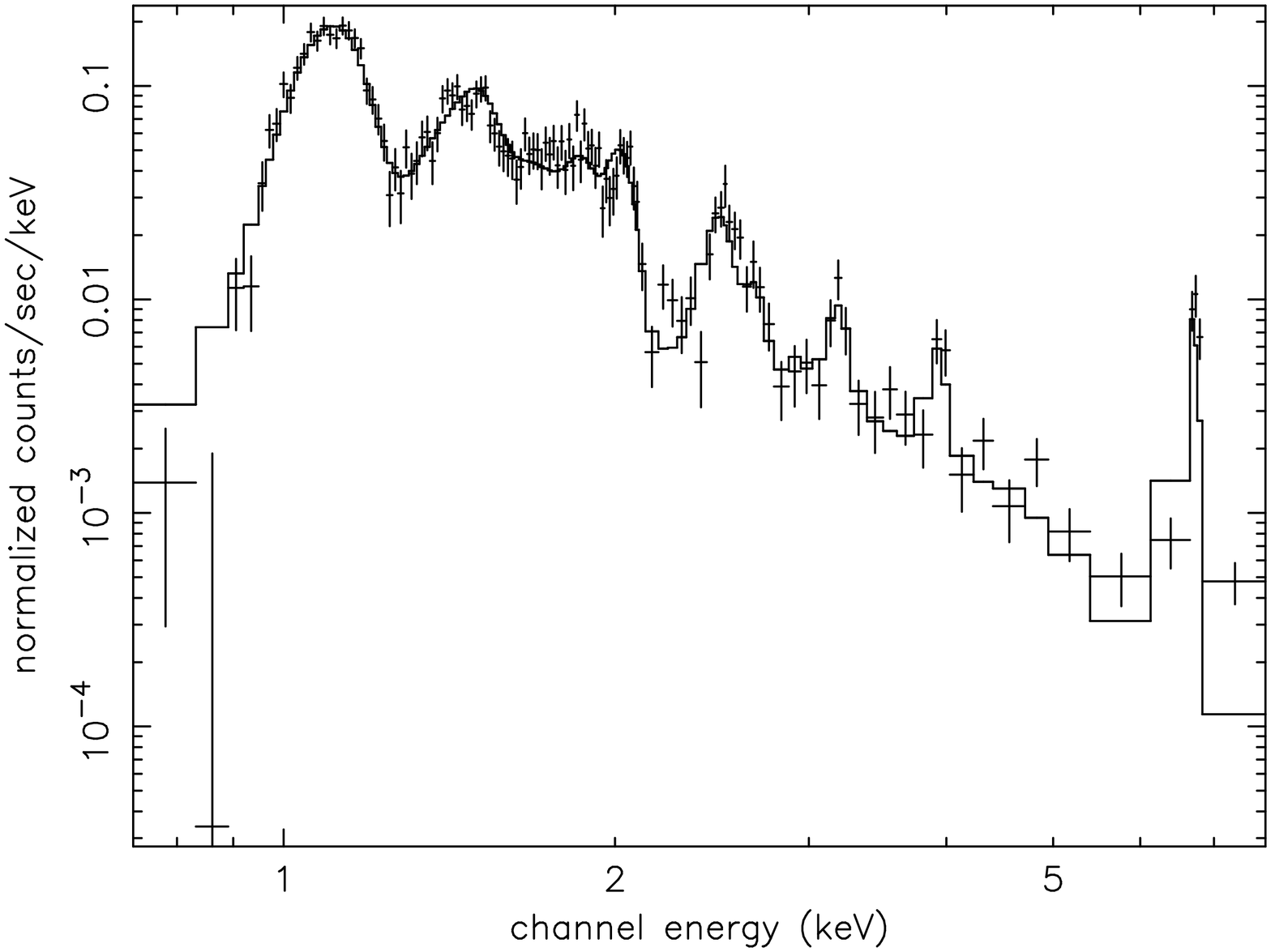}}
\figcaption{Chandra ACIS spectra of a representative Fe-rich knot in
the eastern radial series, fitted with a single temperature, single
ionization age model with O continuum (left) and Si continuum (right)
and the standard off-source background spectrum.  In the lower panel,
the same knot is fitted with a Si continuum model with a local
background.}
\end{figure}

\begin{figure}
\centerline{\includegraphics[scale=0.35,angle=-90]{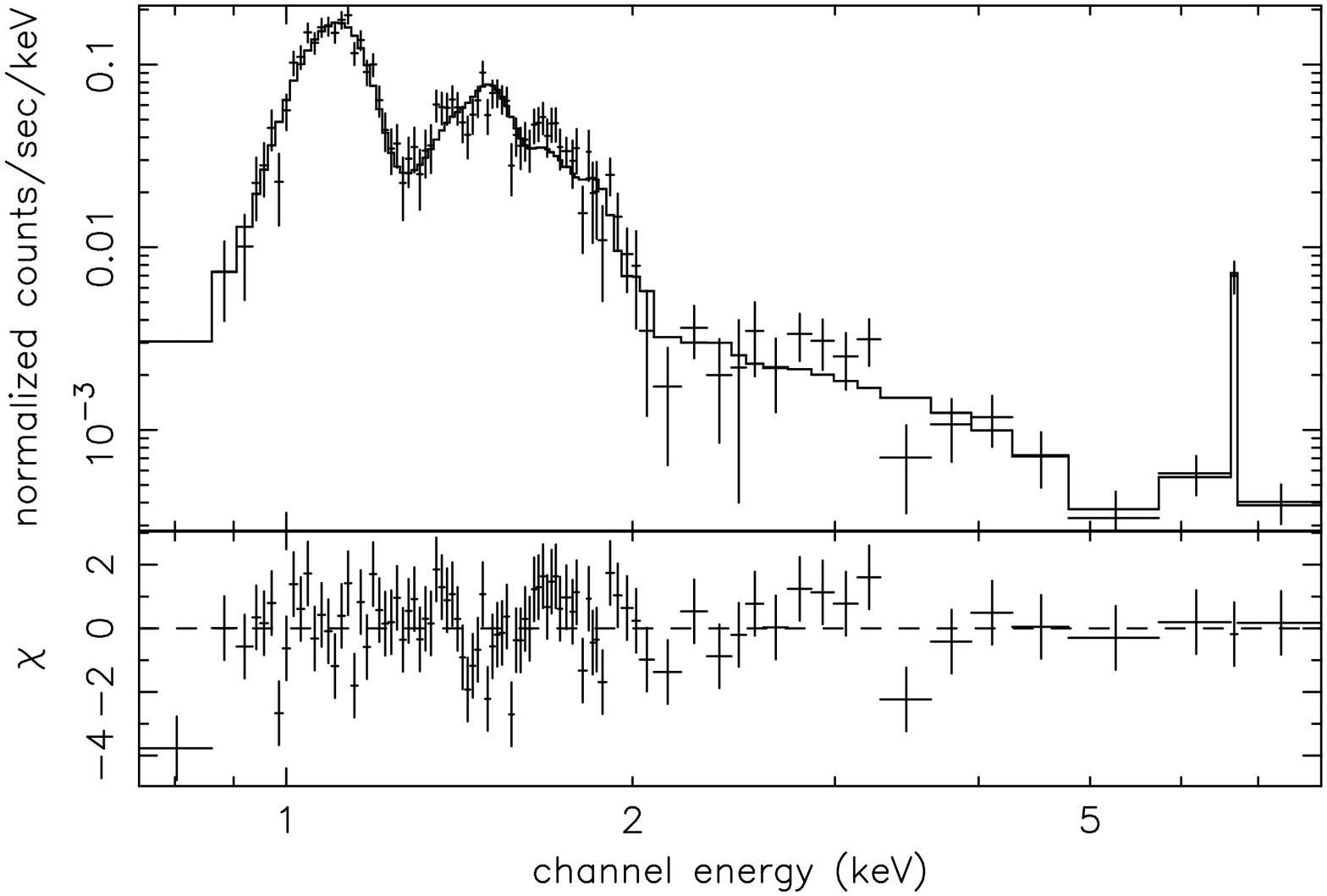}
\includegraphics[scale=0.35,angle=-90]{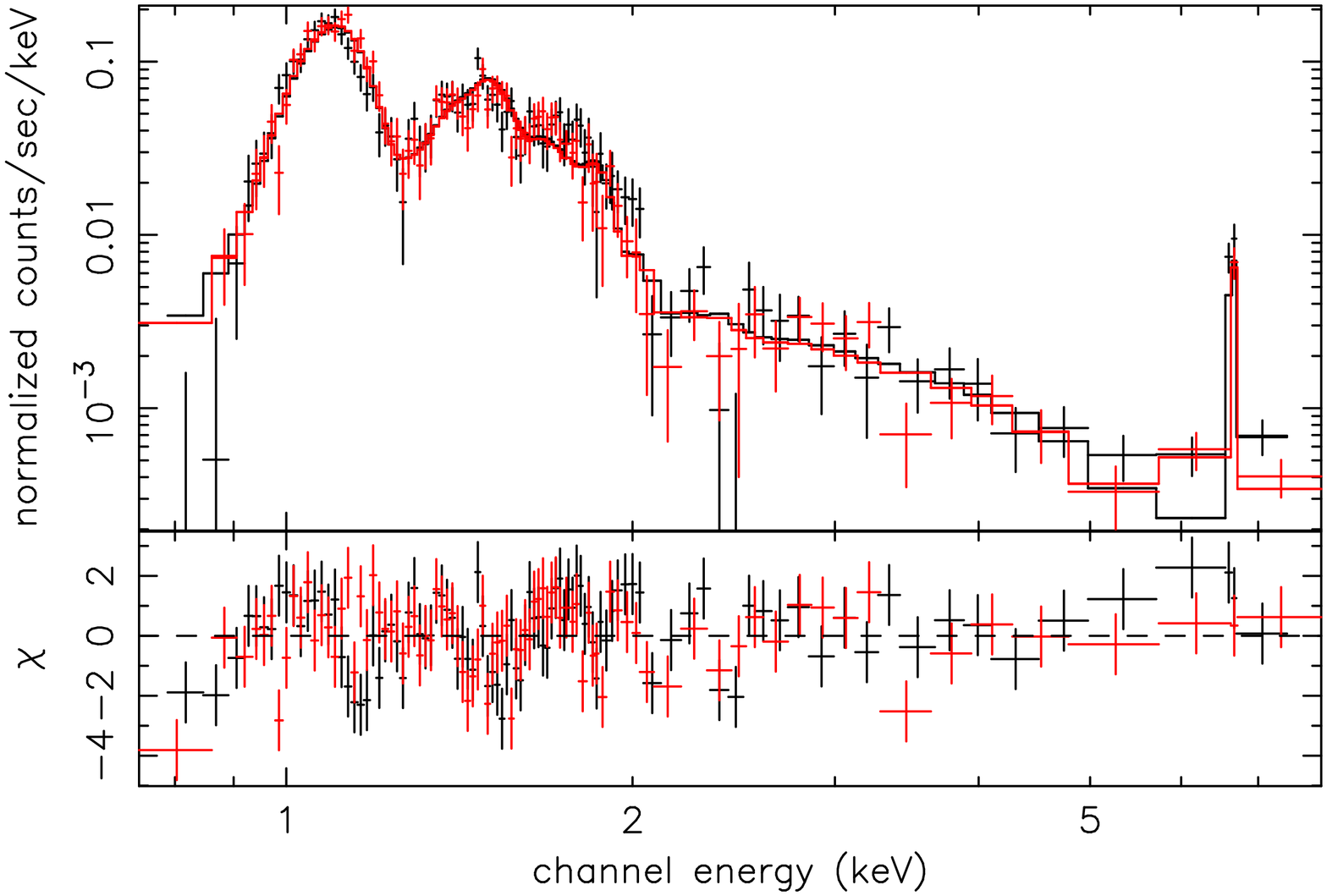}}
\centerline{\includegraphics[scale=0.35,angle=-90]{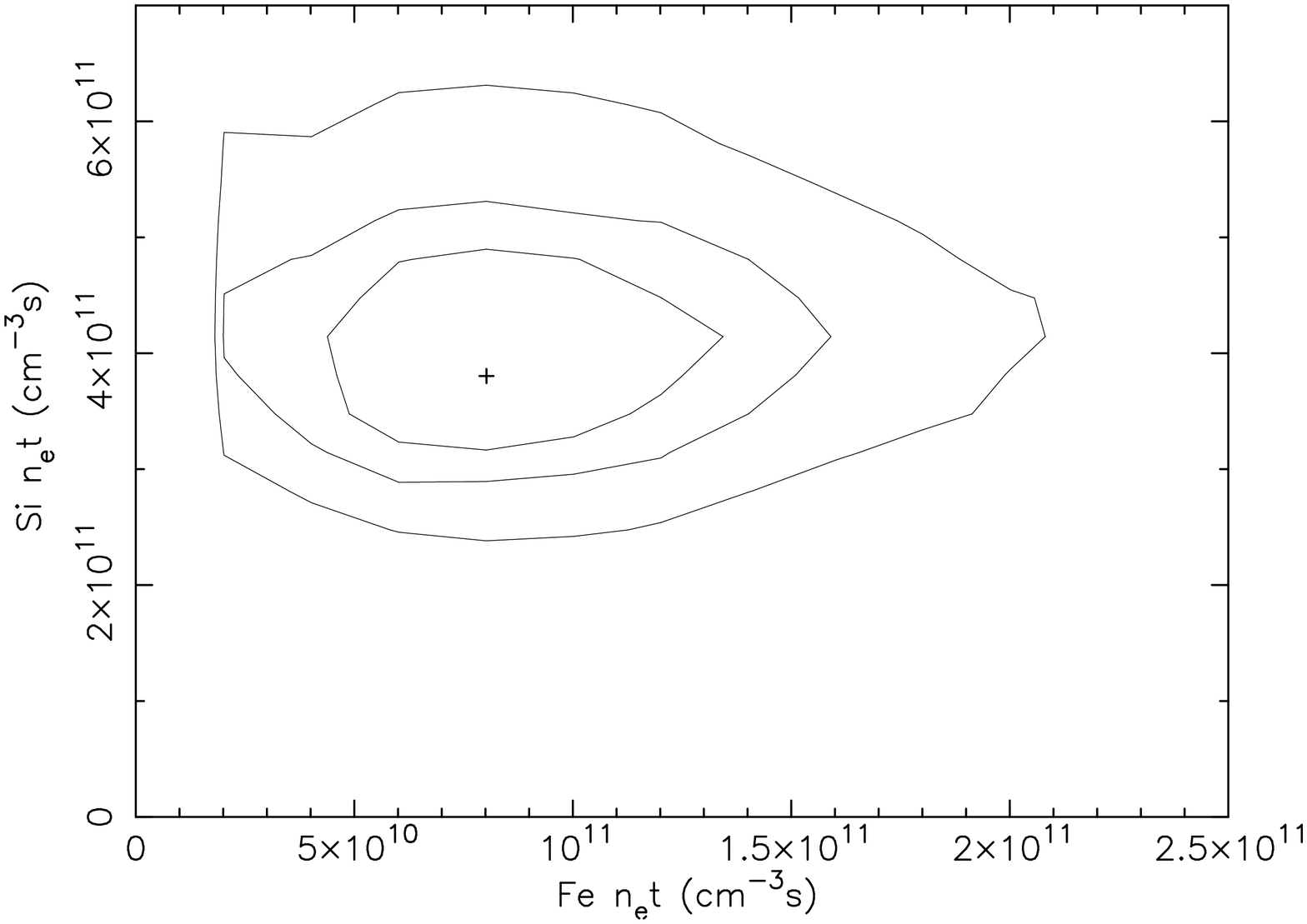}}
\figcaption{(Upper) The 2000 epoch ACIS spectrum of the diffuse Fe
cloud is shown with a single temperature model with separate
ionization ages for Si compared to Fe and Ni (no other elements are
included) in the left panel; the jointly fitted 2000 (red) and 2002
(black) epoch spectra for the Fe cloud are shown in the right
panel. (Lower) Confidence contours for Si and Fe ionization ages
showing $\Delta\chi^2$=2.3,4.6,9.2 for the joint fit shown above. }
\end{figure}

\begin{figure}
\centerline{\includegraphics[scale=1.0]{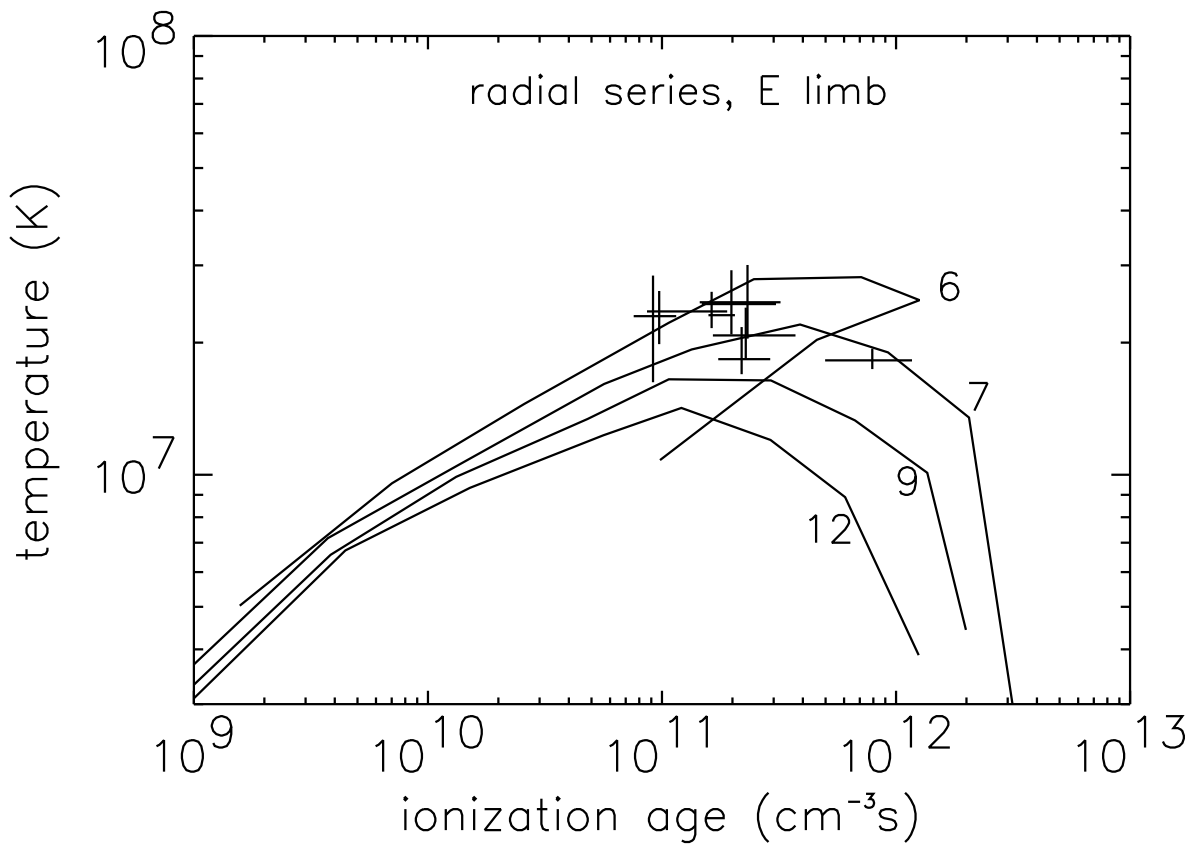}}
\figcaption{Plots of $T_e$ against $n_et$ for varying ejecta envelope
power-law slopes, for a composition O:Si:Fe of 0.83:0.06:0.11 by
mass. Data points from the fits to the ``A'' series of knots are
plotted as crosses, with the size of the cross indicating the fit
uncertainties. The point at highest $n_et$
for $n=6$ corresponds to ejecta at the core-envelope boundary. For higher values
of $n$ this plasma undergoes thermal instability.}
\end{figure}

\begin{figure}
\centerline{\includegraphics[scale=1.0]{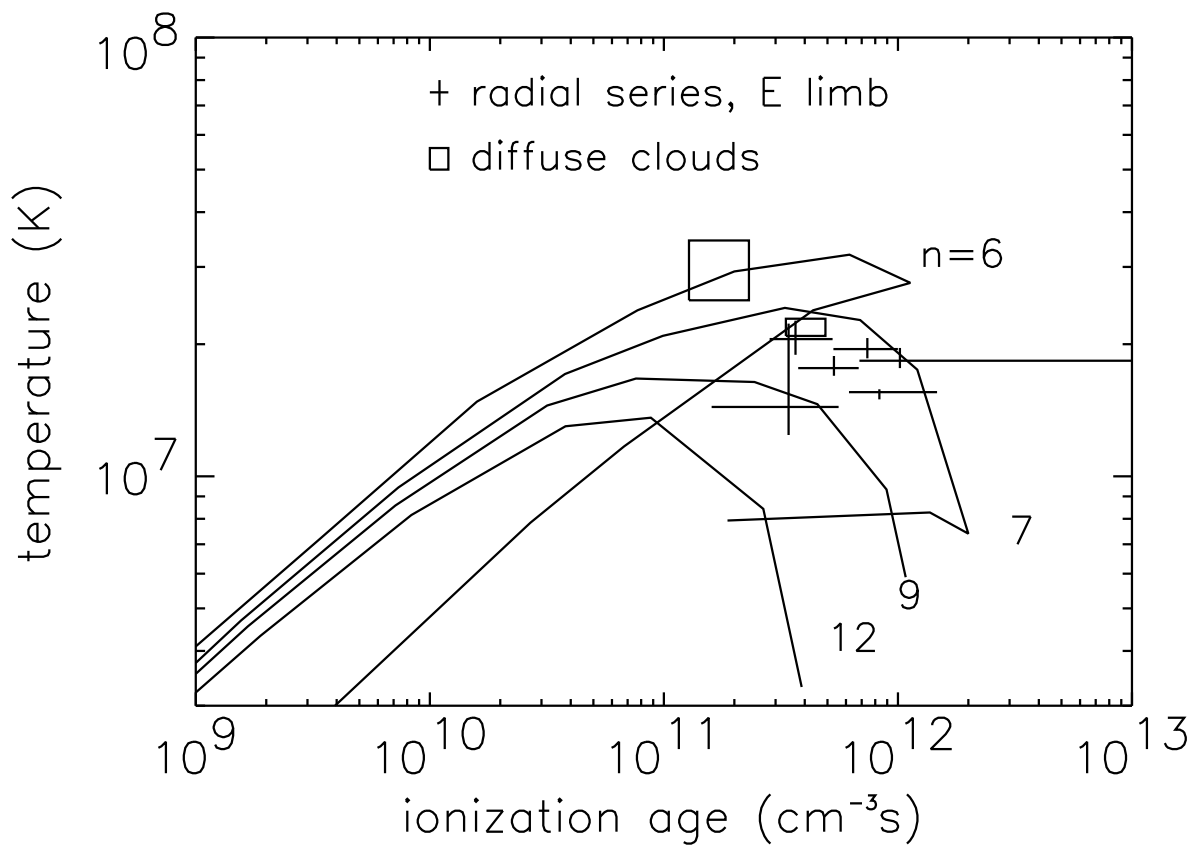}}
\figcaption{Plots of $T_e$ against $n_et$ for varying ejecta envelope
power-law slopes, for a composition Si:Fe of 0.1:0.9 by mass. Data
points from the fits to the ``A'' series of knots are plotted as
crosses, with the size of the cross indicating the fit
uncertainties. The ``diffuse Fe clouds'' are plotted as
boxes, again with size indicating uncertainties.}
\end{figure}

\end{document}